\documentclass[twocolumn,showpacs,10pt,prb,floatfix,amssymb]{revtex4}

\usepackage{graphicx}
\begin{document}

\newlength{\plotwidth}
\setlength{\plotwidth}{8.5cm}

\title{Nanomachining of mesoscopic graphite}

\author{P.~Barthold}
\author{T.~L\"udtke}
\author{R.~J.~Haug}
\affiliation{Institut f\"ur Festk\"orperphysik, Leibniz
Universit\"at Hannover, Appelstr. 2, 30167 Hannover, Germany}
\date{\today}
\begin{abstract}
An atomic force microscope is used to structure a film of
multilayer graphene. The resistance of the sample was measured
\textit{in-situ} during nanomachining narrow trenches. We found a
reversible behavior in the electrical resistance which we
attribute to the movement of dislocations. After several attempts
also permanent changes are observed.
\end{abstract}
\pacs{73.63.-b, 73.22.-f, 73.23.-b, 81.07.-b}
\maketitle

Atomic force microscopes (AFMs) are well known tools for imaging
and for structuring. Besides other lithographic methods
nanomachining with the AFM is a simple, but highly efficient way
to design devices on the sub-micron level. By applying a high
contact force between sample and AFM tip a permanent deformation
of the sample's surface is obtained. Using this method different
materials have been structured e.g.
semiconductors~\cite{Magno_apl_70,schumacher_apl_75,
Regul_APL_81} and metals~\cite{Irmer_apl_73}.\\
Up to now the common technique to structure graphene is by
etching.~\cite{Oezyilmaz_apl_91,stampfer_tunable_nanostructured,russo_condmat_ring}
Graphene has drawn a great deal of attention since the discovery
of free standing single layer graphite (so-called graphene) and
its unique electronic properties.~\cite{Novesolov_science_666,
kim_nature_438,Novoselov_science_315,geim_nature_mat_6} The
motivation for the work presented here was to structure graphene
via nanomachining with an AFM tip. In a first step we structured a
thin film of graphite by nanomachining a trench through the half
width of the sample. Hence the conducting area of the sample is
reduced and thereby a constriction is formed. As we observed
interesting reversible behavior in the resistance in this setup,
in a second step we have cut another sample through the whole
width to see similar effects.\\
The graphite samples used in this study are extracted from natural
graphite~\cite{graphit} by exfoliation \cite{Novoselov_PNAS_102}
on a silicon substrate with a 300~nm SiO$_\mathrm{2}$ layer. The
thereby formed flake has a lateral dimension of a few micrometers
and a thickness of about 10~nm ($\sim$30 atomic layers, assuming a
lattice constant of 0.34~nm). The Ti/Au (9~nm/46~nm) electrodes
are fabricated using standard electron beam lithography. After
bonding the sample it is electrically contacted inside the AFM
allowing \emph{in-situ} measurements at room temperature.
\begin{figure}
\includegraphics[scale=1]{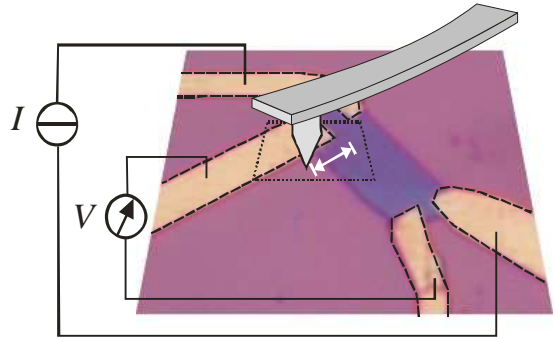}
\caption{\label{fig:setup} Schematic drawing of the setup. The
optical picture shows the graphite flake with four electrodes. A
direct current is driven through the sample via two contacts while
the voltage is measured using the other two electrodes. The AFM
tip moves from left to right and to the left again while a high
contact force is applied. The dashed square marks the region which
is shown in Fig.~\ref{fig:trench}(a) and (b).}
\end{figure}
Figure~\ref{fig:setup} shows the general setup. A direct current
of $I=500$~nA is driven via two contacts through the sample while
the voltage $V$ is measured using the two remaining electrodes.
For the measurements presented here we used an AFM tip that is
coated with polycrystalline diamond on the tip-side. During the
measurements we applied a force of approximately 0.5~$\mu$N. Using
such a high contact force the tip is moved with a velocity of
about 0.5~$\mu$m/s half the way across the graphite flake as
sketched by the white arrows in Fig.~\ref{fig:setup}. The tip
starts its movement left of the flake, moves about 2.2~$\mu$m
through the flake and returns back to its starting position. Thus
the tip scratches the sample in both directions. After five of
those movements a distinct trench
is formed in the graphite film.\\
Figures~\ref{fig:trench}(a) and (b) show two AFM pictures of the
sample before and after nanomachining. A trench of about
2.2~$\mu$m is clearly visible in the graphite flake in
Fig.~\ref{fig:trench}(b).
\begin{figure}
\includegraphics[scale=1]{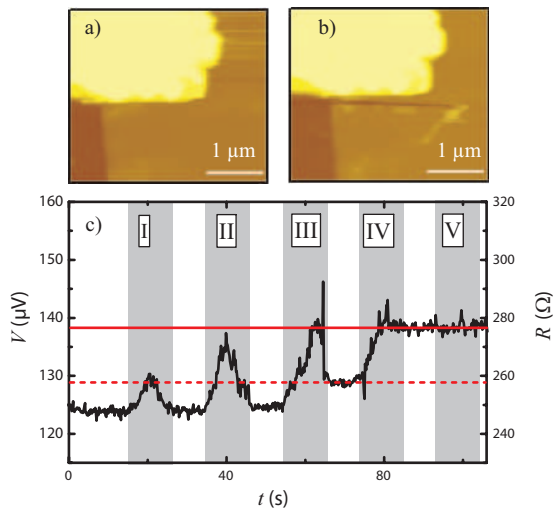}
\caption{\label{fig:trench}Upper part: AFM images of the graphite
flake with a height of about 10~nm. (a) Magnification of the
interesting part as marked in Fig~\ref{fig:setup}. (b) After
nanomachining five times with the AFM tip. A clear trench in the
graphite is visible. Its dimensions are
$w_\mathrm{S}\approx2.2~\mu$m and $l_\mathrm{S}\approx100$~nm. (c)
Time evolution of the resistance of the sample while the AFM tip
applies a force to the graphite. The grey regions indicate when
the tip actually moves on top of the graphite, the roman numerals
count the number of movements. The dashed and solid vertical lines
are guides to the eye to stress the similar resistances of the
sample during different times of structuring.}
\end{figure}
Figure~\ref{fig:trench}(c) demonstrates the time evolution of the
resistance while scratching the graphite film with the AFM tip.
The time period when the tip moves on top of the graphite is
marked grey in Fig.~\ref{fig:trench}(c). At $t=0$~s the resistance
of the sample is about $R\approx248~\Omega$. At $t\approx15$~s
when the tip is moved over the graphite for the first time (I)
with the high contact force the resistance starts to increase. The
resistance reaches its first maximum of $R\approx258~\Omega$ at
$t\approx21$~s which coincides with the reversal point of the AFM
tip movement. The resistance drops again to its original value by
moving the tip back to the original starting position. When the
tip applies a force to the graphite for the second time the
resistance starts to rise again~(II). This time the value rises up
to about 275~$\Omega$. Afterwards the resistance drops to a value
of 248~$\Omega$. As the AFM tip moves over the flake for the third
time (III) the resistance increases to a value of about
277~$\Omega$. Now the resistance decreases to
$R\approx258$~$\Omega$, which is $\Delta R\approx9$~$\Omega$
higher than the resistance in the beginning. The value after the
third tip movement is the same as the maximum obtained during AFM
run I, as indicated by the dashed line in Fig.~\ref{fig:trench}(c)
As the tip moves for the fourth time (IV) over the graphite, the
resistance rises again to a value of about 277~$\Omega$. The same
value is already reached during run II and III. But this time the
resistance does not drop again instead it stays at a value of
$R\approx277~\Omega$. This resistance is kept even when the tip
moves for a fifth time (V) on top of the graphite and stays at
this value afterwards. Thus the resistance of the graphite film
was permanently changed by $\Delta
R\approx29$~$\Omega$ using an AFM tip to structure it.\\
To explain this behavior we consider the following model: While
the AFM tip is moved over the sample dislocations are induced
along the trajectory of the movement of the tip. These
dislocations modify the electronic properties of the sample. Thus
the resistance of the sample rises during scratching. These
dislocations then move to the edge of the sample where we assume
that their influence on the electronic properties of the flake is
only small. Grenall reported dislocation movement in smeared
flakes of natural graphite.~\cite{Grenall_nature_182} As observed
by Williamson dislocations in graphite run parallel to the layer
plane.~\cite{Williamson_proc_royal_257} Mainly they move to the
edge of the flake or to cleavage steps. Hence bonds just destroyed
by the AFM tip along the trajectory of the movement could close
again and the transport properties get back to the original state,
thus the resistance drops again to its original value.\\
As our sample consists of many layers, it seems reasonable to
believe that during the first time the sample is scratched (I)
dislocations are induced only in the few upper layers and during
the second time (II) dislocations are induced in more layers. This
would explain the higher resistance during run~II compared to I.
As the resistance during run~II is close to the value reached at
the end, dislocations seem to be formed in
most of the layers when scratching for the second time.\\
The defects induced during the second time of scratching could
move again to the edge of the sample. Therefore the resistance
drops (between II and III) to its original value. During the third
time of scratching (III) a lasting deformation occurs for the
first time. In a few layers the bonds destroyed by the AFM tip are
not closed again and thereby influence the electronic properties
of the sample permanently. During the fourth run (IV) all layers
are cut through on a 2.2~$\mu$m long path along the sample. Thus
the resistance keeps its value even when it is scratched for the
fifth time. All bonds are destroyed along the trajectory of the
movement of the AFM tip.\\
Using a simple classical resistance network model we find the
following results: Knowing the geometry and measuring the
resistance $R$ of the sample with a four terminal setup we are
able to calculate the specific resistance~$\rho$:
\begin{equation}
\label{eq:spec} \rho=R \cdot \frac{w\cdot h}{l},
\end{equation}
where $l$ is the length in current direction, $w$ the width
orthogonal to the current direction, and $h$ the height of the
sample. The measured values are $R\approx248~\Omega$,
$l\approx7.3~\mu$m, $w\approx5.2~\mu$m, and $h\approx10$~nm. The
resulting specific resistance is $\rho
\approx1.77\cdot10^{-6}~\Omega$m, being comparable to the specific
resistance $\rho\approx1.2\cdot 10^{-6}~\Omega$m reported by
Powell et al. for natural graphite.~\cite{Powell_AIP_142} Now we
estimate the measured resistance during nanomachining using our
simple equivalent circuit by dividing the sample into four parts:
the part above the trench $R_\mathrm{1}$, the area beneath the
trench $R_\mathrm{3}$, the manipulated part $R_\mathrm{S}$, and
the area right to the trench $R_\mathrm{2}$, therefore parallel to
$R_\mathrm{S}$. This leads to an overall resistance given by:
\begin{equation}\label{eq:series}
R_{res}=R_\mathrm{1}+\frac{R_\mathrm{S}\cdot
R_\mathrm{2}}{R_\mathrm{S}+R_\mathrm{2}}+R_\mathrm{3}.
\end{equation}
Nanomachining changes the resistance by $\Delta R \approx
29$~$\Omega$. We now calculate an area of the flake that does not
participate in electrical transport in the end and thereby
explains the change in resistance. We use
$w_\mathrm{S}\approx2.2~\mu$m for the width of the trench. To
determine the other dimension $l_\mathrm{S}^*$ we equate
Eq.~\ref{eq:series} with the resistance in the beginning
($R_{\mathrm{res}}=248~\Omega$) and in the end
($R_\mathrm{res}=277~\Omega$, $R_\mathrm{S}\rightarrow\infty$).
Together with Eq.~\ref{eq:spec} we find
$l_\mathrm{S}^*\approx1.16~\mu$m for the part of the sample which
is nonconducting in the end. This model shows that not only the
trench is insulating in the end, but the area nearby is as well
influenced in its electrical transport properties, as the
calculated $l_\mathrm{S}^*$ is about ten times larger than the
measured $l_\mathrm{S}$ of the trench, clearly showing the limit
of this classical model for a mesoscopic sample.\\
In this model the different resistances observed can be described
by assuming the following dimensions of the resistances in the
equivalent circuit: $l_\mathrm{1}\approx$180~nm,
$w_\mathrm{1}\approx5.2~\mu$m, $l_\mathrm{3}\approx5.9~\mu$m, and
$w_\mathrm{3}\approx5.2~\mu$m for the upper and lower part
$R_\mathrm{1}$ and $R_\mathrm{3}$, respectively. The resistances
in the middle are described by: $w_\mathrm{S}\approx2.2~\mu$m and
$l_\mathrm{S}^*\approx1.16~\mu$m for the left area $R_\mathrm{S}$,
and for the region right to the structured part:
$l_\mathrm{S}^*=l_2\approx1.16~\mu$m and
$w_\mathrm{2}\approx3~\mu$m. For R$_\mathrm{S}\rightarrow\infty$
Eq.~\ref{eq:series} leads to $R\approx277~\Omega$ as measured in
the end. During nanomachining (e.g. during run I) we have to
assume a height of $h_S=5~$nm for $R_\mathrm{S}$ instead of
$h_S=10$~nm at the beginning and $h_S=0$~nm at the end.\\
\begin{figure}
\includegraphics[scale=1]{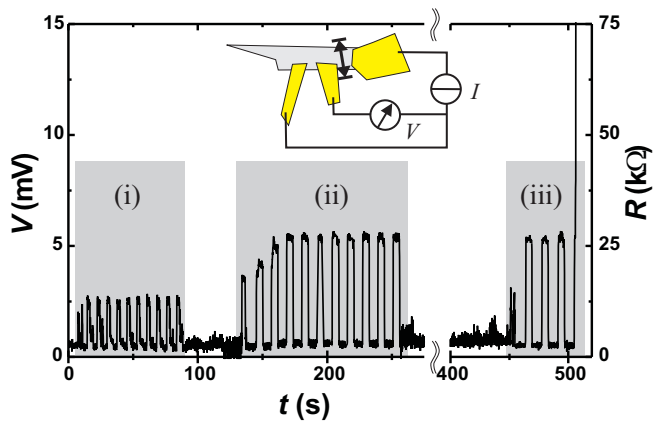}
\caption{\label{fig:different_sample}Measurement of sample 2. The
inset shows a sketch of the setup. The three gold electrodes and
the graphite flake are shown. The arrow indicates the trajectory
of the tip moved over the whole sample. The \emph{in-situ}
measurement of voltage is shown on the left axis and the
corresponding resistance on the right axis. In each interval (i),
(ii), and (iii) a different contact force is applied to the
sample. At $t\approx510$~s the resistance rises to the measurable
limit. Hence the graphite is cut through.}
\end{figure}
In Fig.~\ref{fig:different_sample} the measured resistance for
another sample is shown. In this case we use a three terminal
setup as sketched in the inset to drive a current through the
device and to measure the voltage. The sample is structured in
three different intervals, during each interval the tip is moved
eleven times through the sample with different forces:
$<0.1~\mu$N, about $1~\mu$N, and approximately $2~\mu$N. In the
last interval during the fourth time of moving the tip, the
resistance reaches the detectable limit. Hence the graphite film
is cut through. The dramatic change in the resistance, when the
tip moves over the sample, is clearly visible. During interval (i)
the resistance rises by $\Delta R\approx10$~k$\Omega$. In return
the resistance drops again, when no force is applied. During
interval (ii) the resistance increases by values up to $\Delta
R\approx24$~k$\Omega$. This reversible effect is very pronounced
in Fig.~\ref{fig:different_sample}. The same resistance is reached
many times by nanomachining even after a longer pause between
interval (ii) and (iii).\\
We again attribute this observed reversible change in the
resistance to induced and moving dislocations as before. As this
setup is a three terminal device we have to take a contact
resistance into account. By this we again find quite good
qualitative agreement assuming a classical series of
resistances according to the geometry of the sample.\\
In conclusion, we have shown \textit{in-situ} measurements of the
resistance of mesoscopic graphite being nanomachined with a
diamond coated AFM tip. In doing so we find a reversible change in
the electrical resistance. We attribute this effect to induced
dislocations that lead to an increased resistance. At room
temperature these dislocations can easily move to the edges of the
graphite flake leading to reversible resistance changes. After
processing a sample with the AFM tip a couple of times the
resistance changes permanently, i.e. bonds inside the graphite are
broken permanently.
\newpage

\bibliographystyle{prsty}

\newpage

\end{document}